\documentclass[conference,comsoc]{IEEEtran}

\usepackage{float}
\usepackage{graphicx}
\usepackage{clrscode}
\usepackage{stfloats}
\usepackage{paralist}
\usepackage{cite}
\usepackage{comment}
\usepackage{graphicx}
\usepackage{amsmath, amssymb, mathrsfs, amsfonts, amsthm}
\usepackage{epsfig, epstopdf}
\usepackage[tight,footnotesize]{subfigure}
\usepackage{algorithm}
\usepackage{color}
\usepackage[usenames,dvipsnames,svgnames,table]{xcolor}
\usepackage{bbold}
   
\usepackage{chemarrow}                          
\usepackage{algorithmic}

\usepackage{tikz}
\usetikzlibrary{arrows,automata}
\usepackage{mathtools}
\usepackage{enumitem}
\usepackage{algorithmic}
\usepackage{hyperref}

\newtheorem{lemma}{Lemma}

\def\msP{\mathsf{P}}

\newcommand{\HH}{\mathcal{H}_0}
\newcommand{\HHH}{\mathcal{H}_1}

\newcommand{\Pf}{ \text{P}_{\text{fa}} }
\newcommand{\Pd}{ \text{P}_{\text{d}} }
\newcommand{\Pm}{ \text{P}_{\text{m}} }

\newcommand{\Pfl}{ \text{P}_{\text{fa}}^{\text{T-S}} }
\newcommand{\Pdl}{ \text{P}_{\text{d},k}^{\text{T-S}} }

\newcommand{\Pflf}{ \text{P}_{\text{fa},k}^{\text{S-F}} }
\newcommand{\Pdlf}{ \text{P}_{\text{d},k}^{\text{S-F}} }
\newcommand{\E}{\mathbb{E}}
\newcommand{\Po}{\mathsf{Poisson}}
\newcommand{\Ex}{\mathsf{exp}}
\newcommand{\J}{\zeta}

\newcommand{\g}{\mathsf{g}}
             
\newcommand{\argmax}{\arg\!\max}
\newcommand{\K}{k_{\text{d}}}
\newcommand{\Db}{D}
\newcommand{\Kf}{k_{\text{f}}}
\newcommand{\Kb}{k_{\text{b}}}

\newcommand{\rr}{\Vert \mathbf{x}_T-\mathbf{x}_k \Vert}
\newcommand{\hh}{\mathsf{H}}

\allowdisplaybreaks

\begin{document}

\title{Advanced Target Detection via Molecular Communication\vspace{-4mm}}

\author{Reza Mosayebi$^\dagger$, Wayan~Wicke$^{\ddagger}$, Vahid~Jamali$^{\ddagger}$, Arman~Ahmadzadeh$^{\ddagger}$,\\ Robert~Schober$^{\ddagger}$, and~Masoumeh~Nasiri-Kenari$^\dagger$ \\
$^\dagger$Sharif University of Technology, Tehran, Iran; $^{\ddagger}$University of Erlangen-Nuremberg, Erlangen, Germany \vspace{-5mm} }

\maketitle 
\vspace{-15mm}
\begin{abstract} 
In this paper, we consider target detection in suspicious tissue via diffusive molecular communications (MCs). If a target is present, it continuously and with a constant rate secretes molecules of a specific type, so-called \emph{biomarkers}, into the medium, which are symptomatic for the presence of the target. 
Detection of these biomarkers is challenging since due to the diffusion and degradation, the biomarkers are only detectable in the vicinity of the target. In addition, the exact location of the target within the tissue is not known. In this paper, we propose to distribute several reactive nanosensors (NSs) across the tissue such that at least some of them are expected to come in contact with biomarkers, which cause them to become activated.
Upon activation, an NS releases a certain number of molecules of a secondary type into the medium to alert a fusion center (FC), where the final decision regarding the presence of the target is made.
In particular, we consider a composite hypothesis testing framework where it is assumed that the location of the target and the biomarker secretion rate are unknown, whereas the locations of the NSs are known. We derive the uniformly most powerful (UMP) test for the detection at the NSs. For the final decision at the FC, we show that the UMP test does not exist. Hence, we derive a \textit{genie-aided} detector as an upper bound on performance. We then propose two sub-optimal detectors and evaluate their performance via simulations.\vspace*{0mm}
\end{abstract}

\vspace*{-0mm}
\section{Introduction}\label{sec:intro}
\vspace*{-0mm}
During the past years, diffusion-based molecular communication (MC) systems have received significant attention as a candidate for the design of bio-inspired nanonetworks, because of their size scale, biocompatibility, and low energy consumption \cite{Nakano}. Bio-inspired nanonetworks have many potential applications, especially in healthcare and environmental monitoring \cite{Nakano}.

One of the fundamental challenges in healthcare monitoring is the problem of early detection of signs of an anomaly in the body, like the presence of a tumor \cite{Mishra}, which we refer to as \emph{target detection} \cite{Nakano2}. In particular, for tumor detection, direct detection of the tumor cells themselves is difficult since their size is small and their locations are unknown. Instead, a significant body of research has been devoted to detecting protein molecules, referred to as \emph{biomarkers}, that are secreted by the tumor cells into blood vessels and tissue \cite{Wuab, Stanstna}. 
However, detection of these biomarkers is also challenging since due to diffusion and degradation, the biomarkers might be detectable only in the vicinity of the target. In addition, the exact location of the target within the tissue and the secretion rate of the biomarkers are in general not known. Detection is possible if a sensor with the ability to detect these biomarkers is placed in the vicinity of the target.
Due to recent advances in nanotechnology, one interesting approach to detect biomarkers is to employ engineered nanosensors (NSs) \cite{Wuab, Chen}. 


Target detection in MC systems is different from target detection in wireless sensor networks because of the signal-dependent noise in the MC channel and the possibility of reactions between molecules.
In the MC literature, the problem of anomaly detection was considered in \cite{Felicetti, Nakano_Let, Reza_blood, Ghavami_Detection, Lahouti_Detection, Reza_TNB}. 
In \cite{Felicetti, Nakano_Let, Reza_blood} the use of mobile nanosensors (MNSs) is proposed to detect the presence of anomaly in the vasculature.  
In particular, in \cite{Felicetti} and \cite{Nakano_Let}, it is assumed that MNSs move through the vasculature and gather at the target location by binding to the target. In our recent work \cite{Reza_blood}, anomaly detection using MNSs is proposed where the MNSs are activated if they come in contact with biomarkers. The MNSs move through the vasculature and are then collected by a fusion center (FC), which decides on the presence of anomaly.
In \cite{Ghavami_Detection, Lahouti_Detection, Reza_TNB} employing fixed NSs is proposed for target detection in body tissue. In particular, in \cite{Ghavami_Detection}, \cite{Lahouti_Detection}, the channel between the NSs and the FC is assumed to be an additive white Gaussian noise channel; while in \cite{Reza_TNB} a Poisson signal-dependent noise channel is considered.

In this paper, similar to \cite{Ghavami_Detection, Lahouti_Detection, Reza_TNB}, we assume that multiple NSs are placed on the surface of a suspicious tissue, which we refer to as the \emph{surveillance area}, along with an FC. However, we adopt a more realistic receiver model compared to \cite{Ghavami_Detection, Lahouti_Detection, Reza_TNB} for the NSs and the FC, respectively, namely a general reactive receiver model \cite{Arman}.
Similar to \cite{Reza_blood}, we assume that a target continuously releases biomarkers into the tissue, including the surveillance area. If a biomarker reaches an NS, it may react with the receptors on the surface of the NS and thus activate them. If the number of activated receptors of an NS exceeds a threshold, it will secrete a certain number of molecules of a secondary unique type that is detectable by the reactive FC into the environment.
Unlike \cite{Reza_TNB}, we make the realistic assumption that both the location of the possible target and its biomarker secretion rate are unknown. However, the locations of the NSs are assumed to be known. 
The main contributions of this paper can be summarized as follows:
\begin{enumerate}
\item We derive an analytical expression for the probability mass function (PMF) of the number of activated receptors on the surface of an NS which is a function of the target location and the \emph{continuous} biomarker secretion rate. We validate the result with particle-based simulation of Brownian motion and the reactive receiver. 
\item Next, we derive the optimal hard detection scheme for the NSs and show that it corresponds to a uniformly most powerful (UMP) test. The UMP test is a test that, without knowledge of unknown parameters, performs equal to the optimal Neyman-Pearson detector \cite{Kay} that knows the parameters. In our case, these parameters are the location of the target and the biomarker secretion rate. 
\item Finally, we develop a composite hypothesis testing framework for the FC, where the location of the possible target and the secretion rate of the biomarkers are unknown. We then derive a \emph{genie-aided} detector (GAD), which provides a performance upper bound for any realizable detector at the FC. Furthermore, we propose two sub-optimal detectors for practical detection. The performance of the proposed detectors is evaluated via Monte Carlo simulation.
\end{enumerate}

The rest of this paper is organized as follows. Section~II introduces the system model. In Section~III, we derive the PMF of the number of activated receptors and the optimal detector for the NSs. In Section~IV, we introduce the GAD and propose two sub-optimal detectors for the FC. We evaluate the performance of the proposed detectors via simulations in Section~V, and conclude the paper in Section~VI.

\section{System Model And Preliminaries} \label{sec:model}

\subsection{System Model}\label{sec:IIA}
\begin{figure}[!t]
	\centering
	\includegraphics[scale=0.38]{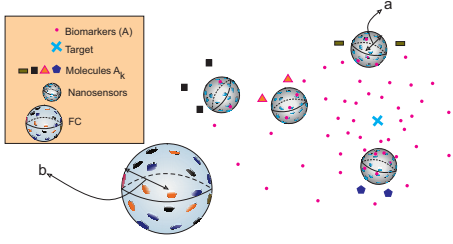}\vspace*{-1 mm}
	\caption{Schematic diagram of the considered system with $K=4$ sensors.}\vspace*{-2 mm}
	\label{Fig:1}
\end{figure}

We consider an unbounded three-dimensional (3-D) environment with constant temperature and viscosity, a possible target located at position $\mathbf{x}_T$, $K$ identical spherical reactive NSs with radius $a$ located at positions $\mathbf{x}_k, k\in \mathcal{K}\triangleq \{1, 2, ..., K\}$, and a spherical FC located at $\mathbf{x}_F$ with radius $b$, as depicted in Fig. \ref{Fig:1}.

When the target is present in the surveillance area, we assume that it continuously secretes biomarkers, which are denoted as type $A$ molecules, at position $\mathbf{x}_T$ into the environment with secretion rate $\mu$ [biomarkers$\cdot$s$^{-1}$]. 
We denote the presence (abnormality) and absence (normality) of the target by hypotheses $\HHH$ and $\HH$, respectively. 
The secreted biomarkers independently diffuse in the environment with constant diffusion coefficient $\Db$ and may reach an NS. We furthermore assume that the secreted biomarkers can degrade at a rate of $\K$ [s$^{-1}$] via a first-order degradation reaction of the form
\begin{align}
 A \overset{\K}{\rightarrow} \oslash,
\end{align}
where $\oslash$ is a species of molecules that is not recognized by the NSs nor the FC.

%
We assume that each NS has $M$ receptor proteins on its surface, which we refer to as $B$ molecules and are modeled as disks with radius $r_{\text{d}}$. Biomarkers that come in contact with an NS may reversibly react with the $B$ molecules on the surface of the NS and activate them via a second-order reversible reaction as follows 
\begin{align}
 A+B \overset{\Kf}{\underset{\Kb}{\rightleftarrows}} C,
\end{align}
where $\Kf$ and $\Kb$ are forward and backward reaction rates in molecule$^{-1}$m$^3$s$^{-1}$  and s$^{-1}$, respectively, and one generated molecule of type $C$ represents one activated receptor. 
We define $\msP \left(t;\mathbf{x}_T, \mathbf{x}_k \right)$ as the probability that a given biomarker secreted by the target at time $t=0$ and at location $\mathbf{x}_T$ activates a receptor of the $k$-th NS centered at location $\mathbf{x}_k$ at time $t$. We also assume that there are other sources of type $A$ molecules. Theses additional molecules are regarded as environmental noise.
For the number of activated receptors at the $k$-th NS, which constitutes the received signal, we define random variable (RV) $Y_k$ and its realization $y_k$. Based on $y_k$, the $k$-th NS makes a local hard decision denoted by $c_k$ regarding the presence of the target. If the NS decides that the target is present, it relays this information to the FC by a one-shot instantaneous emission of $N$ secondary molecules of type $A_k$ at time $T_1$ with diffusion coefficient $D_k$; otherwise, the NS does not emit any molecule. For simplicity, we assume that the emission time $T_1$ for all NSs is identical.

The molecules released by the NSs may reach the FC, which can perform more complex operations than the NSs. Furthermore, the FC may be connected to an outside computer which may perform computationally expensive processing tasks if needed. We assume that the FC is reactive with respect to all molecules of type $A_k, k \in \mathcal{K}$, and has $M'$ receptors of type $B_k, k \in \mathcal{K}$, on its surface, where each receptor is modeled by a disk with radius $r_{\text{F}}$. 
For the secondary molecules, we also assume first-order degradation reactions in the channel and second-order reversible reactions at the FC, i.e.,
\begin{align}
 A_k \overset{k_{d,k}}{\rightarrow} \oslash,~
 A_k+B_k \overset{k_{\text{f},k}}{\underset{k_{\text{b},k}}{\rightleftarrows}} C_k,~~ k \in \mathcal{K},\vspace*{-10 mm}
\end{align}
where $k_{\text{f},k}$, $k_{\text{b},k}$, and $k_{\text{d},k}$ are the corresponding forward, backward, and degradation rates, respectively. 
We also define $\msP \left( T_2-T_1;\mathbf{x}_k, \mathbf{x}_F \right)$ as the probability that a given molecule of type $A_k$, emitted at time $T_1$ at $\mathbf{x}_k$, activates a receptor of type $B_k$ at time $T_2$ at the surface of the FC centered at $\mathbf{x}_F$.
In addition, we assume that the FC is able to count the numbers of molecules of type $C_k, k \in \mathcal{K}$, at time $T_2$, which corresponds to the signal received from the $k$-th NS and is modeled by RV $Z_k$ and its realization $z_k$. We assume a source of environmental noise for all type $C_k$ molecules.
At the FC, for the tractability of the analysis and to reduce the complexity of the decision rule, we assume that first a hard decision, denoted by $d_k$, is made regarding the relayed message from the $k$-th NS by comparing $z_k$ with a specific threshold. Then, based on all $K$ hard decisions, which are collected in vector $\mathbf{d}=[d_1, d_2, \cdots, d_K]^T$, the FC decides on the presence of the target. 

\vspace*{-0mm}
\subsection{Preliminaries}\label{sec:pre}
\vspace*{-0mm}
In this subsection, we briefly review the reactive receiver model in \cite{Arman}, as we adopted this model for both the NSs and the FC. In particular, we derive the receptor activation probability at the $k$-th NS, $\msP \left(t;\mathbf{x}_T, \mathbf{x}_k \right)$, as a function of the parameters of the channel between the target and the $k$-th NS. 
 
Using \cite{Arman} and assuming that the NSs are placed sufficiently far from each other such that their received signals do not influence each other, the received signals of different NSs can be assumed to be independent. Therefore, we obtain the following expression for $\msP (t;\mathbf{x}_k, \mathbf{x}_T)$ \cite[Eq. (29)]{Arman}
\begin{align}\label{eq:one}
&\msP (t;\mathbf{x}_T, \mathbf{x}_k) = \frac{\Kf~\Ex\left( -\K t \right)}{4\pi\sqrt{D}a \rr}  \Biggl \lbrace{ \frac {\alpha \text {W} \left ({ \frac {\rr - a}{\sqrt { 4 \Db t}}, \alpha \sqrt {t} }\right )}{(\gamma - \alpha )(\alpha - \beta )} } \notag \\ 
&~{\! +\, \frac {\beta \text {W} \!\left ({ \frac {\rr - a}{\sqrt {4 \Db t}}, \beta \sqrt {t} \!}\right )}{(\beta - \gamma )(\alpha - \beta )} + \! \frac {\gamma \text {W} \!\left ({ \frac {\rr - a}{\sqrt {4 \Db t}}, \gamma \sqrt {t} \!}\right )}{(\beta - \gamma )(\gamma - \alpha )} }\Biggr \rbrace,
\end{align}
where 
$\Vert \cdot\Vert$ is the $\ell_2$ norm, $\text{W}(n,m) = \Ex(2nm+m^2)\mathsf{erfc}(n+m)$, $\mathsf{erfc}(\cdot)$ denotes the complementary error function, and $\alpha$, $\beta$, and $\gamma$ are the solutions of the following system of non-linear equations
\begin{equation}\label{eq:alpha}
 \begin{cases} \alpha + \beta + \gamma = \left ({ 1 + \frac {k^{\star }_{\text{f}}}{4 \pi a \Db} }\right ) \frac {\sqrt { \Db}}{a}, \\ \alpha \gamma + \beta \gamma + \alpha \beta = \Kb - \K, \\ \alpha \beta \gamma = \Kb \frac {\sqrt { \Db}}{a} - \K \left ({ 1 + \frac {k^{\star }_{\text{f}}}{4 \pi a \Db} }\right ) \frac {\sqrt { \Db}}{a}. 
 \end{cases} 
\end{equation} 
We note that $\alpha, \beta,$ and $\gamma$ can be complex numbers.
In (\ref{eq:alpha}), $k^{\star }_{\text{f}}$ is given by 
\begin{align}
{k}^{\star }_{\text{f}} = \frac {4\pi\Db\Kf\hspace*{0.5mm}\varphi }{ \Kf a (1 - \varphi ) + 4 \pi \Db}, 
\end{align}
where $\varphi$ is the same for all NSs and is given by
\begin{equation} 
\varphi \!=\! \frac {M r_{d}^{2} ( \Kf a + 4 \pi \Db)}{a^{2}(1 \!-\! \lambda )(\pi r_{d} \Kf\!+\! 16 \pi \Db) + M r_{d}^{2} ( \Kf a \!+\! 4 \pi \Db)},\qquad 
\end{equation}
and $\lambda = M\pi r_d^2/(4\pi a^2)$.

Similarly, the receptor activation probability for the $k$-th receptor type at the FC,
$\msP \left( T_2-T_1;\mathbf{x}_k, \mathbf{x}_F \right)$, can be obtained from (\ref{eq:one}) by substituting $M$, $r_{\text{d}}$, $a$, $D$, $\Kf$, $\Kb$, and $\K$ with $M'$, $r_{\text{F}}$, $b$, $D_k$, $k_{\text{f},k}$, $k_{\text{b},k}$, and $k_{\text{d},k}$, respectively, and also solving (\ref{eq:alpha}) with the parameters of the channel between the $k$-th NS and the FC.\vspace*{-0mm}
%
%
%
\section{Detector Design at Nanosensors}\label{sec:local}
\vspace*{-0mm}
In this section, we first derive the steady-state PMF of the received signal at the $k$-th NS. Then, based on the derived PMF, we derive the optimal local decision rule at the $k$-th NS. \vspace*{-0mm}

\subsection{PMF of the Received Signal at the NSs}
\vspace*{-0mm}
In this subsection, first, we derive the average value of the received signal, i.e., the mean number of activated receptors at the $k$-th NS. Then, given this mean, we derive the PMF of the received signal.
By using (\ref{eq:one}) and considering the fact that the target is secreting biomarkers at a constant rate of $\mu$, the average value of the received signal at the $k$-th NS at time $t$ can be obtained by integrating $\msP (t;\mathbf{x}_T, \mathbf{x}_k)$ over time, i.e.,
\begin{align}\label{mean}
\int_{0}^{t} \mu \hspace*{0.5mm} \msP (\tau;\mathbf{x}_T, \mathbf{x}_k) \mathsf{d}\tau.
\end{align}
The asymptotic mean value of the received signal at the $k$-th NS, denoted by $m_k$, is obtained for  $t \rightarrow \infty$, and is given in Lemma \ref{lemma:1}.

\begin{lemma}\label{lemma:1}
The steady-state average value of the received signal at the $k$-th NS is given by
\begin{align}\label{eq:steady}
&m_k = \mu \hspace*{0.5mm} \g(\mathbf{x}_T,\mathbf{x}_k) \triangleq \frac{\mu \hspace*{0.5mm} \Kf~\Ex\left( \left(-\rr+a \right)\sqrt{\frac{\K}{\Db}} \right)}{4\pi\sqrt{D}\hspace*{0.5mm}a \rr} \notag \\
  & \times\Biggl \lbrace{\hspace*{-0.5mm} \frac {\alpha  }{(\gamma - \alpha )(\alpha - \beta )(\alpha \sqrt{\K}+\K)} \hspace*{-0.5mm}+ \hspace*{-0.5mm}\frac {\beta  }{(\beta - \gamma )(\alpha - \beta )(\beta \sqrt{\K}+\K)} } \notag \\ 
& ~~{\!+ \ \frac {\gamma  }{(\beta - \gamma )(\gamma - \alpha )(\gamma \sqrt{\K}+\K)} }\Biggr \rbrace .
\end{align}
\end{lemma}
\begin{IEEEproof} 
Due to the space limitation, we only provide a sketch of the proof. In particular, by substituting (\ref{eq:one}) in (\ref{mean}), taking the limit $t\rightarrow\infty$, and using the following integral \cite[Eq. (4.3.34)]{Edward}
\begin{align}\label{eq10}
& \int_{x=0}^{\infty} \mathsf{erfc}\left(ax+ \frac{b}{x}\right) \Ex \left( -c^2x^2\right) x \hspace*{0.5mm}\mathsf{d} x =  \frac{1}{2} (a^2+c^2)^{-\frac{1}{2}} \nonumber \\
&\times \left( a+\sqrt{a^2+c^2} \right)^{-1} \Ex \left( -2b(a+\sqrt{a^2+c^2}) \right),
\end{align}
which holds if $\mathcal{R}(b) >0$ and $\mathcal{R}(a^2+c^2)>0$, where $\mathcal{R}(\cdot)$ is the real part operator, we arrive at (\ref{eq:steady}). 
\end{IEEEproof} 
In Section~\ref{sec:numeric}, we show that for a finite $t$, the average value of the received signal closely approaches the asymptotic value.

In the following, we calculate the PMF of the received signal at the NSs, which we use for the subsequent analysis. When the target continuously secretes biomarkers, since the release time instances of the biomarkers are different, the PMF of the received signal at the $k$-th NS follows a general \emph{Poisson binomial} distribution. 
Although the Poisson binomial distribution is cumbersome to work with, it can often be approximated by a Poisson distribution when the number of trials (secreted biomarkers) is large and the success probability $\msP (t;\mathbf{x}_T, \mathbf{x}_k)$ is small, cf. \cite{Binomial}. Since these conditions are met in typical MC environments, we approximate the received signal at the $k$-th NS by a Poisson RV. The accuracy of this approximation is evaluated in Section~\ref{sec:numeric}. Furthermore, we also model the additive and independent environmental noise by a Poisson distribution \cite{Yilmaz_Poiss,Reza_Receiver} with mean $\J_0$, which is present under both hypotheses $\HH$ and $\HHH$. 
Hence, the received signal at the $k$-th NS is modeled by
\begin{align}\label{eq:local-dist}
 Y_k \sim \left\{ 
\begin{array}{l l}
\Po(\J_0), ~~~ &\text{under}~\HH,\\
\Po(\J_0+\mu \hspace*{0.5mm}\g(\mathbf{x}_T,\mathbf{x}_k)), ~~~ &\text{under}~\HHH,\\
\end{array} \right.
\end{align}
where $\Po(x)$ denotes the Poisson distribution with mean $x$ and $\g (\cdot, \cdot)$ is defined in \eqref{eq:steady}.
Since $\mu$ and $\mathbf{x}_T$ are not known, we obtain the following composite hypothesis testing problem
\begin{align}\label{hypoo}
 \left\{ 
\begin{array}{l l}
\HH: ~~~ &\text{if}~\mu = 0,\\
\HHH: ~~~ &\text{if}~\mu > 0,~~~~ \mathbf{x}_T~(\text{nuisance}). \\
\end{array} \right.
\end{align}
We note that the discriminator parameter between the two hypotheses is $\mu$; while $\mathbf{x}_T$ is a nuisance parameter that only exists for hypothesis $\HHH$.\vspace*{-0mm}

\subsection{Optimal Local Detector at the NSs}\label{sec:IIIb}
\vspace*{-0mm}
In this subsection, our goal is to design the optimal local detector based on the received signal $y_k$ at the $k$-th NS that maximizes the local detection probability subject to a pre-assigned upper bound $\omega_1$ on the local false alarm probability at the NS, i.e.,
\begin{align}\label{criteria}
 &\max_{\text{local detectors}} \Pdl,~ \text{subject to}~ \Pfl \leq \omega_1,
\end{align}
where $\Pfl$ and $\Pdl$ are respectively the local false alarm and the detection probabilities for the link between the target and the $k$-th NS. We note that $\Pfl$ only depends on the decision threshold and $\J_0$, while $\Pdl$ depends on the decsion threshold, $\J_0$, and the unknown parameters $\mu$ and $\mathbf{x}_T$. If the location of the target and its biomarker secretion rate were known at the NSs, the optimal detector for (\ref{criteria}) would be the Neyman-Pearson detector \cite{Kay}, which compares the local log-likelihood ratio (LLR) with the maximum threshold that ensures $\Pfl \leq \omega_1$. Denoting the local LLR of the $k$-th NS as $\lambda_k$, we obtain
\begin{align}\label{lambda}
\lambda_k &\triangleq \log \left( \frac{ \msP\left( y_k \big|\HHH \right)}{\msP\left( y_k \big| \HH \right)} \right) \nonumber \\
&= \log \left( \frac{\mu \hspace*{0.5mm} \g(\mathbf{x}_T,\mathbf{x}_k)+\J_0}{\J_0} \right)y_k - \mu \hspace*{0.5mm}\g(\mathbf{x}_T,\mathbf{x}_k),  
\end{align}
where $\msP(\cdot)$ denotes probability. From (\ref{lambda}), we obtain that independent of the value of the unknown parameters $\mu>0$ and $\mathbf{x}_T$, the local LLR $\lambda_k$ is a monotonic increasing function of $y_k$. Therefore, by using the Karlin-Rubin theorem \cite{Karlin}, we arrive at the following UMP test: 
\begin{align}\label{opt:loc}
\text{c}_{k} = \left\{ 
\begin{array}{l l}
0, ~~~ &\text{if}~~ y_k \leq \tau_1,\\
1, ~~~ &\text{if}~~ y_k > \tau_1,\\
\end{array} \right.
\end{align}
where $\tau_1$ is the decision threshold at the NSs and is the same for all NSs. Clearly, since $\Pfl$ in (\ref{opt:loc}) is a decreasing function of $\tau_1$, we obtain $\tau_1=\{ \max \tau\hspace*{-1.2mm}: \Pfl(\tau)-\omega_1 \leq 0\}$. Therefore, instead of comparing $\lambda_k$, which we cannot evaluate due to the unknown parameters $\mu$ and $\mathbf{x}_T$, we can directly compare observation $y_k$ with $\tau_1$, which yields the same performance as the optimal Neyman-Pearson detector that employs (\ref{lambda}) as decision variable and compares it with a corresponding maximum threshold such that $\Pfl \leq \omega_1$. 
Given the proposed optimal local detector in (\ref{opt:loc}), we can evaluate the local false alarm probability as follows
\begin{align}\label{fal}
\Pfl \hspace{-0.1cm}&=\hspace{-0.05cm} \msP\left( c_k = 1 \big|\HH \right) \hspace{-0.05cm} =  \msP\left( y_k > \tau_1 \big|\HH \right) \nonumber \\
 &= \sum_{i=\tau_1+1}^{\infty} \hspace{-0.1cm} \frac{\Ex(-\J_0)(\J_0)^i}{i!} \hspace{-0.05cm} \triangleq \hspace{-0.05cm} \hh\left( \tau_1, \J_0 \right)\hspace{-0.05cm},
\end{align}
which is a decreasing function of $\tau_1$.
Similarly, we can evaluate the local detection probability of the $k$-th NS as
\begin{align}\label{dal}
\Pdl = \msP\left( c_k = 1 \big|\HHH \right) = \hh \left( \tau_1, \J_0+\mu \hspace*{0.5mm} \g(\mathbf{x}_T,\mathbf{x}_k) \right).
\end{align}


\section{Detector Design at the FC} \label{sec:FC}
As mentioned in Section~\ref{sec:model}, the FC is a reactive receiver sensitive to all $A_k, k \in \mathcal{K}$, molecules. 
Employing the same approach as in Section~\ref{sec:IIIb} for deriving the local decision rule of the NSs, we arrive at the following initial hard decision rule at the FC for detection of the signal sent by the $k$-th NS
\begin{align}\label{opt:loc2}
\text{d}_{k} = \left\{ 
\begin{array}{l l}
0, ~~~ &\text{if}~~ z_k \leq \tau_2,\\
1, ~~~ &\text{if}~~ z_k > \tau_2,\\
\end{array} \right.
\end{align}
where $z_k$ is the received signal from the $k$-th NS, i.e., the number of type $C_k$ molecules produced at time $T_2$ at the FC, and $\tau_2$ is the threshold that the FC employs to detect the signal received from the NSs. For simplicity, $\tau_2$ is assumed to be identical for all NSs. Since each NS secretes the secondary molecules instantaneously at time $T_1$, $Z_k$ is distributed as
\begin{align}\label{eq:local-dist2}
 Z_k \sim \left\{ 
\begin{array}{l l}
\Po(\J_k), ~ &\text{if}~c_k=0,\\
\Po(\J_k+N \msP (T_2-T_1;\mathbf{x}_F, \mathbf{x}_k) ), ~ &\text{if}~c_k=1,\\
\end{array} \right. 
\end{align}
where $\J_k, k \in \mathcal{K}$, is the average number of environmental noise molecules of type $C_k, k \in \mathcal{K}$, and $N$ is the number of molecules of type $A_k$ released by the $k$-th NS if $c_k=1$.  
Similar to (\ref{fal}) and (\ref{dal}), we can derive the false alarm and detection probabilities for the hard decision rule in (\ref{opt:loc2}) for the link between the $k$-th NS and the FC, which we denote by $\Pflf$ and $\Pdlf$, respectively. We note that since the locations of the NSs are assumed to be known at the FC, both $\Pflf$ and $\Pdlf$ are known by the FC for all $k \in \mathcal{K}$.

To model the hypothesis test at the FC, we express $d_k$ in (\ref{opt:loc2}) in terms of the hypotheses $\HH$ and $\HHH$. To this end, by considering (\ref{opt:loc}) and (\ref{opt:loc2}) we arrive at a binary non-symmetric channel between the target and the FC with the following transition probabilities
\begin{align}\label{eqopt:trans}
\msP \hspace{-0.05cm}\left(d_k=1\right) = \left\{ 
\begin{array}{l l}
\Pdlf~\Pfl + \Pflf (1-\Pfl) \triangleq \rho_{0,k},&\text{under}~\HH, \\
\Pdlf~\Pdl + \Pflf (1-\Pdl) \triangleq \rho_{1,k},&\text{under}~\HHH, \\
\end{array} \right.
\end{align}
where $\rho_{1,k}$ is a function of the unknown parameters $\mu$ and $\mathbf{x}_T$, i.e., we can also write $\rho_{1,k}(\mathbf{x}_T,\mu)$.

At the FC, the goal is to design the optimal detector (based on hard decision vector $\mathbf{d}$) that maximizes the global detection probability denoted by $\Pd$ subject to a pre-assigned upper bound $\omega_2$ on the global false alarm probability denoted by $\Pf$, and given $\tau_1$ and $\tau_2$. That is,
\begin{align}\label{criteria2}
 \max_{\text{detectors}} \Pd,~ \text{subject to}~ \Pf \leq \omega_2,~ \text{and given}~ \tau_1~\text{and}~ \tau_2.
\end{align}
Now, similar to Section \ref{sec:IIIb}, if $\mathbf{x}_T$ and $\mu$ were known at the FC, the optimal detector for (\ref{criteria2}) would compare the total LLR of $\mathbf{d}$ with the maximum threshold such that $\Pf \leq \omega_2$. The total LLR can be obtained as follows
\begin{align}\label{LLR}
\overline{\mathsf{LLR}} &= \log \left( \frac{ \msP\left( \mathbf{d} \big|\HHH \right)}{\msP\left( \mathbf{d} \big|\HH \right)} \right) = \sum_{k=1}^{K}\mathsf{LLR}_k \\
&\triangleq \sum_{k=1}^{K} \biggl \lbrace{  \log \left( \frac{\rho_{1,k}}{\rho_{0,k}}\right)d_k + \log \left( \frac{1-\rho_{1,k}}{1-\rho_{0,k}}\right) (1-d_k)    \biggr \rbrace}. \notag
\end{align}
Since $\rho_{1,k}$ is a function of unknown parameters $\mu$ and $\mathbf{x}_T$, we have a similar composite hypothesis testing problem as in (\ref{hypoo}), and thus cannot directly employ (\ref{LLR}). Nevertheless, as a benchmark scheme, we use (\ref{LLR}) and assume $\mathbf{x}_T$ and $\mu$ are known. This detector is referred to as GAD and yields an upper bound on the achievable performance of any practical detector at the FC that does not know $\mathbf{x}_T$ and $\mu$. 

Unlike (\ref{lambda}), (\ref{LLR}) may not be a monotonic function of $d_k$. Therefore, we cannot use the Karlin-Rubin theorem to change the structure of the detector in (\ref{LLR}) such that it does not require knowledge of $\mu$ and $\mathbf{x}_T$. 
In addition, since the nuisance parameter $\mathbf{x}_T$ apprears only under hypothesis $\HHH$, we cannot directly use many of the detectors proposed in the literature for composite hypothesis testing, such as the locally optimum detector (LOD) or Rao and Wald tests \cite{Kay}.
Instead, in the following subsections, we derive two (generally) sub-optimal decision rules for the FC.\vspace*{-0mm}

\subsection{Generalized-Likelihood Ratio Test}\label{sec:GLRT}
\vspace*{-0mm}
A common approach for the composite hypothesis testing is the generalized-likelihood ratio test (G-LRT). The G-LRT decision variable can be expressed as \cite{Kay}:
\begin{align}\label{GLR}
\mathsf{T}_{\text{G-LRT}} 
&= 2 \sum_{k=1}^{K} 
\Biggl \lbrace{  \mathsf{log}\left( \frac{\rho_{1,k}\left( \widehat{\mathbf{x}}_T, \widehat{\mu} \right)}{\rho_{0,k}} \right)d_k  } \notag \\ 
&{\! ~~~+ \mathsf{log}\left( \frac{1-\rho_{1,k}\left( \widehat{\mathbf{x}}_T, \widehat{\mu} \right)}{1-\rho_{0,k}} \right) (1-d_k) } \Biggr \rbrace,   
\end{align}
where $\widehat{\mathbf{x}}_T$, and $\widehat{\mu}$ denote the ML estimates of $\mathbf{x}_T$ and $\mu$ under hypothesis $\HHH$, i.e.,
\begin{align}\label{muxhat}
\left(\widehat{\mathbf{x}}_T, \widehat{\mu} \right) \hspace*{-1mm} &= \hspace*{-1mm} \argmax_{\mathbf{x}_T, \mu} \msP\left( \mathbf{d} \big|\HHH, \mu, \mathbf{x}_T  \right) \nonumber \\
&= \argmax_{\mathbf{x}_T, \mu} \biggl \lbrace{  \mathsf{log} \left( \rho_{1,k}\left( \widehat{\mathbf{x}}_T, \widehat{\mu} \right) \right)d_k } \nonumber \\
&~{\! +\  \mathsf{log} \hspace*{-0.7mm}\left(1- \rho_{1,k}\left( \widehat{\mathbf{x}}_T, \widehat{\mu} \right) \right) (1-d_k)} \biggl \rbrace.
\end{align}
To perform the G-LRT, the decision variable in (\ref{GLR}) is compared with the maximum threshold such that $\Pf \leq \omega_2$, which we denote by $\tau_3$.
Since we cannot analytically solve \eqref{muxhat} for $\widehat{\mu}$ and $\widehat{\mathbf{x}}_T$, we find $\widehat{\mu}$ and $\widehat{\mathbf{x}}_T$ numerically, e.g. via a grid search, cf. Section \ref{sec:numeric}. As a result, the complexity of the G-LRT is high since the search has to be performed with respect to both $\mathbf{x}_T$ and $\mu$. Hence, to reduce the complexity of the final decision rule at the FC, in the following subsection, we derive another detector which is less complex.\vspace{-0mm}

\subsection{Generalized-Locally Optimum Detector }\label{sec:GLOD}
\vspace*{-0mm}
A different approach for the case that the unknown parameters are only present under hypothesis $\HHH$ is the detector proposed in \cite{Davies}. In the following, we refer to this detector as \emph{generalized-locally optimum detector} (G-LOD), to underline the use of an LOD\footnote{Following \cite{Kay}, it can be proven that the performance of the LOD is close to the Neyman-Pearson detector if $\mu$ is very small.} in the decision rule. 
The decision variable of the G-LOD can be written as
\begin{align} \label{GLOD}
\mathsf{T}_{\text{G-LOD}} = \max_{\mathbf{x}_T} \frac{\frac{\partial~ \mathsf{log}\left( \msP\left( \mathbf{d} \big|\HHH, \mu, \mathbf{x}_T  \right) \right) }{\partial \mu} }{\sqrt{\mathsf{I}\left(\mathbf{x}_T, \mu = 0 \right)}}\Bigg|_{\mu = 0},
\end{align}
where $\mathsf{I}\left(\mathbf{x}_T, \mu \right)$ denotes the Fisher information, i.e.,
\begin{align}\label{FI}
\mathsf{I}\left(\mathbf{x}_T, \mu \right) = \E \hspace*{0.5mm} {\left( \frac{\partial~ \mathsf{log}\left( \msP\left( \mathbf{d} \big|\HHH, \mu, \mathbf{x}_T  \right) \right) }{\partial \mu} \right)^2 },
\end{align}
and $\E(\cdot)$ is the expectation operator.
In the following, first we derive the derivative in the numerator of (\ref{GLOD}) before we use it to evaluate (\ref{FI}). The derivative can be written as
\begin{align}\label{partial}
\frac{\partial~ \mathsf{log}\left( \msP\left( \mathbf{d} \big|\HHH, \mu, \mathbf{x}_T  \right) \right) }{\partial \mu} = \sum_{k=1}^{K}
\frac{\frac{\partial\rho_{1,k}}{\partial \mu}}{\rho_{1,k}}d_k - \frac{\frac{\partial\rho_{1,k}}{\partial \mu}}{1-\rho_{1,k}}(1-d_k). 
\end{align}
To proceed, we need $({\partial\rho_{1,k}})/({\partial \mu})$ which is given in Lemma \ref{lemma:2}.
\begin{lemma} \label{lemma:2}
The derivative of the PMF of decision vector $\mathbf{d}$ with respect to $\mu$ under hypothesis $\HHH$ is given by
\begin{align}\label{eq:lemma2}
\frac{\partial\rho_{1,k}}{\partial \mu} &= \frac{ \Ex \left( -\mu \hspace*{0.5mm} \g( \mathbf{x}_T, \mathbf{x}_k ) -\J_0\right) \left(\mu \hspace*{0.5mm}\g( \mathbf{x}_T, \mathbf{x}_k ) +\J_0\right)^{\tau_1}  }{\tau_1!} \notag \\
&~~\times \g\left( \mathbf{x}_T, \mathbf{x}_k\right) \left( \Pdlf-\Pflf \right). 
\end{align}
\end{lemma}
\begin{IEEEproof} 
By replacing (\ref{fal}) and (\ref{dal}) in (\ref{partial}), we obtain
\begin{align}\label{eq:midDeriv}
&\frac{\partial\rho_{1,k}}{\partial \mu} =  \frac{\partial}{\partial \mu} \left( \mathsf{H}\left(\tau_1,\mu \hspace*{0.5mm}\g( \mathbf{x}_T, \mathbf{x}_k) +\J_0  \right) \right) \left( \Pdlf-\Pflf\right), 
\end{align}
where
\begin{align}\label{eq:Hderiv}
&\frac{\partial}{\partial \mu} \left( \mathsf{H}\left(\tau_1,\mu \hspace*{0.5mm}\g( \mathbf{x}_T, \mathbf{x}_k) +\J_0  \right) \right)  \notag \\
&=  \sum_{k=\tau_1+1}^{\infty}\frac{\partial}{\partial \mu} \left( \frac{ \Ex \left( -\mu \hspace*{0.5mm} \g( \mathbf{x}_T, \mathbf{x}_k ) -\J_0\right) \left(\mu \hspace*{0.5mm} \g( \mathbf{x}_T, \mathbf{x}_k ) +\J_0\right)^{k}  }{k!} \right)  \nonumber \\
&= \sum_{k=\tau_1+1}^{\infty} \frac{1}{k!}   \biggl \lbrace{  -\g( \mathbf{x}_T, \mathbf{x}_k ) \Ex ( -\mu \hspace*{0.5mm}\g( \mathbf{x}_T, \mathbf{x}_k ) -\J_0) }   \notag \\
&{\!  ~~\times(\mu \hspace*{0.5mm}\g( \mathbf{x}_T, \mathbf{x}_k ) +\J_0)^{k} } \biggr \rbrace +   
\sum_{k=\tau_1+1}^{\infty} \frac{1}{k!}   \biggl \lbrace{  k\hspace*{0.5mm}\g( \mathbf{x}_T, \mathbf{x}_k )  }   \notag \\
&{\! ~~\times \Ex ( -\mu \hspace*{0.5mm}\g( \mathbf{x}_T, \mathbf{x}_k ) -\J_0)(\mu \hspace*{0.5mm}\g( \mathbf{x}_T, \mathbf{x}_k ) +\J_0)^{k-1} } \biggr \rbrace \notag \\
&= \frac{ \Ex \left( -\mu \hspace*{0.5mm}\g( \mathbf{x}_T, \mathbf{x}_k ) -\J_0\right) \left(\mu \hspace*{0.5mm}\g( \mathbf{x}_T, \mathbf{x}_k ) +\J_0\right)^{\tau_1}  }{\tau_1!}  \g\left( \mathbf{x}_T, \mathbf{x}_k\right).
\end{align}
By substituting (\ref{eq:Hderiv}) in (\ref{eq:midDeriv}), we arrive at (\ref{eq:lemma2}). This completes the proof.
\end{IEEEproof} 

Using similar algebraic calculations, we obtain the following expression for the Fisher information in (\ref{FI})
\begin{align}\label{eq:FILast}
\mathsf{I}\left(\mathbf{x}_T, \mu \right) \hspace*{-1mm} &= \hspace*{-1mm} \left( \frac{ \Ex \left( -\mu \hspace*{0.5mm}\g( \mathbf{x}_T, \mathbf{x}_k ) -\J_0\right) \left(\mu \hspace*{0.5mm}\g( \mathbf{x}_T, \mathbf{x}_k ) +\J_0\right)^{\tau_1}  }{\tau_1!} \right)^2 \notag \\
&~~\times  \frac{\left( \g\left( \mathbf{x}_T, \mathbf{x}_k\right) \left( \Pdlf-\Pflf \right) \right)^2}{\rho_{1,k}(1-\rho_{1,k})}. 
\end{align} 
Now, by plugging (\ref{partial}), (\ref{eq:lemma2}), and (\ref{eq:FILast}) into (\ref{GLOD}), we obtain the following decision variable for the G-LOD
\begin{align}\label{GLODF}
\mathsf{T}_{\text{G-LOD}} = \max_{\mathbf{x}_T} \frac{\sum_{k=1}^{K} \vartheta_k \left( \frac{d_k}{\rho_{0,k}}-\frac{1-d_k}{1-\rho_{0,k}} \right) }{\sqrt{\sum_{k=1}^{K}\frac{\vartheta_k^2}{\rho_{0,k}(1-\rho_{0,k})}}},
\end{align}
where
\begin{align}
\vartheta_k \triangleq  \frac{ \g\left( \mathbf{x}_T, \mathbf{x}_k\right) \Ex \left( -\J_0\right) (\J_0)^{\tau_1}  }{\tau_1!} \left( \Pdlf-\Pflf \right). 
\end{align}
The G-LOD makes the final decision on the presence of the target by comparing (\ref{GLODF}) with threshold $\tau_3$. Thereby, the complexity of the G-LOD is lower than that of the G-LRT, since the maximization in (\ref{GLODF}) is only with respect to $\mathbf{x}_T$.

\vspace*{-0mm}
\section{Numerical Results} \label{sec:numeric}
In this section, first, we validate the results derived for the mean and the distribution of the number of activated receptors (received signal) at an NS via the particle-based simulator developed in \cite{Arman}. Then, we consider a sample network consisting of several NSs and evaluate the performance of the proposed detectors by plotting the global probability of missed detection $\Pm=(1-\Pd)$ and the global probability of false alarm $(\Pf)$. Table~\ref{table:2} summarizes the system parameters that were used for all simulations, unless stated otherwise. Here, ``mol'' is used for the abbreviation of ``molecule''.
\begin{table}
\renewcommand{\arraystretch}{1.2}
\centering
\caption{List of Important Simulation Parameters For the Individual Network Components \cite{Arman} and the Network Topology.} \vspace*{-2 mm}
\begin{tabular}{|l|l|l|l|} 
\hline
Param. & Value &  Param. & Value \\ 
\hline
$\Db, D_k$  &  $5 \times 10^{3}\mu$m$^2$s$^{-1}$ & $\mu$  &  $10^{3}$s$^{-1}$  \\
\hline
$\K$   & $10^{-4}$s$^{-1}$  & $k_{\text{d},k}$ & $5 \times 10^{-7}$s$^{-1}$ \\
\hline
$\Kf$  & $1.2\times 10^{4}\mu$m$^3$s$^{-1}$mol$^{-1}$  & $k_{\text{f},k}$\hspace*{-7mm} & $3.7\times 10^{4}\mu$m$^3$s$^{-1}$mol$^{-1}$\hspace*{-2mm}  \\ 
\hline
$\Kb$  & $1.5\times 10^{-4}~$s$^{-1}$  & $k_{\text{b},k}$ &  $5\times 10^{-6}~$s$^{-1}$ \\  
\hline
$M$, $M'$ & $5.12\times 10^{3}$ & $\{a$, $b\}$ & $\{0.5, 1\}~\mu$m \\  
\hline
$r_{\text{d}}$ & $7\times 10^{-3}\mu$m & $r_{\text{F}}$& $1.4\times 10^{-2}\mu$m \\  
\hline
$\J_0$ & $10$ & $\J_k$& $5$ \\ 
\hline
$\tau_1$ & $16$ & $\tau_2$& $9$ \\ 
\hline
$\mathbf{x}_T$  &  $(10,10,0)~\mu$m & $\mathbf{x}_F$ & $(-30,-30,0)~\mu$m  \\
\hline
$T_1$  &  $10~$ms & $T_2$ & $15~$ms  \\
\hline
\end{tabular} \vspace*{-2 mm}
\label{table:2}
\end{table} 

\begin{figure}[!t]
	\centering
	\includegraphics[scale=0.25]{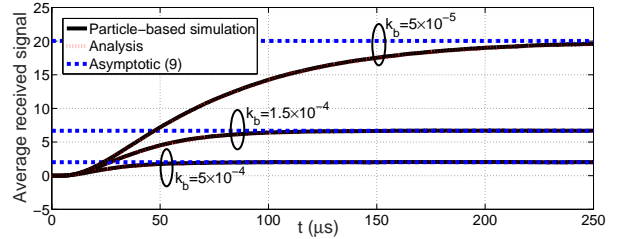}\vspace*{-2 mm}  
	\caption{Average received signal at an NS centered at $(1,0,0)~\mu$m as a function of time for different $\Kb$.}
	\vspace*{-1 mm}
	\label{Fig::1}
\end{figure}

\begin{figure}[!t]
	\centering
	\includegraphics[scale=0.25]{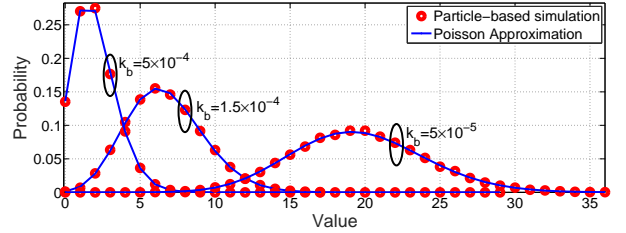}\vspace*{-2 mm}
	\caption{Poisson approximation and histogram, obtained from particle-based simulation for the received signal at an NS centered at $(1,0,0)\mu$m for different $\Kb$.}\vspace*{-2 mm}
    \label{Fig::2}
\end{figure}

Fig. \ref{Fig::1} depicts the average received signal at an NS versus time for system parameters $\mathbf{x}_k=(1,0,0)~\mu$m, $\mathbf{x}_T=(0,0,0)~\mu$m, and $\mu=10^3$.
The particle-based simulation results in Figs. \ref{Fig::1} and \ref{Fig::2} were averaged over $2\times 10^4$ independent realizations of the channel with a simulation step size of $5\times 10^{-2}\mu$s. 
Since there are three reaction rates $\Kf, \Kb$, and $\K$, due to the space limit, we only present results for the case when $\Kb$ is altered. In Fig. \ref{Fig::1}, we show three sets of curves, where the analytical curves are obtained by numerically evaluating (\ref{mean}) for each $t$ and the asymptotic curves are obtained from (\ref{eq:steady}).
We observe that the analytical and simulation results are in excellent agreement. In addition,
for all considered values of $\Kb$, the average received signal reaches its asymptotic value before $250\mu$s. 
We also observe that when $\Kb$ increases, the asymptotic mean number of activated receptors decreases. This is due to the fact that as $\Kb$ increases, the rate of the backward reaction increases which reduces the number of activated receptors at a given time.

Fig. \ref{Fig::2} shows the histograms for the received signal at an NS for the same parameters as in Fig. \ref{Fig::1} at time $t=T_2=15~$ms. In Fig. \ref{Fig::2}, we also show the Poisson PMF approximations for the received signal at the NS in (\ref{eq:local-dist}) for $\J_0=0$. As can be observed, the histogram of the received signal at the NS is very well approximated by the Poisson PMF for all considered scenarios. Hence, this result confirms the accuracy of the proposed approximation of the received signal at the NS by a Poisson PMF with the mean in (\ref{eq:steady}). Similar observations have been made for the signals received from the NSs at the FC.

\begin{figure}[!t]
	\centering
	\includegraphics[scale=0.30]{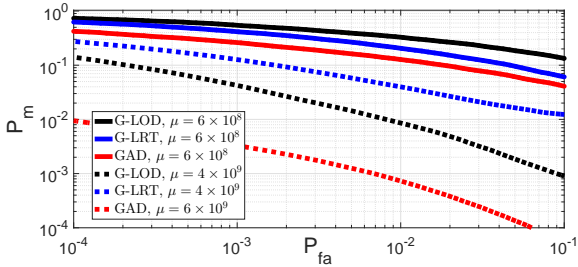}
	\vspace*{-2 mm} 
	\caption{Global probability of missed detection versus global probability of false alarm for $N=10^7$, and $\mu \in \{6\times 10^{8}, 4\times 10^{9}\}$.}\vspace*{-2 mm}
\label{Fig::3}
\end{figure}

\begin{figure}[!t]
	\centering
	\includegraphics[scale=0.30]{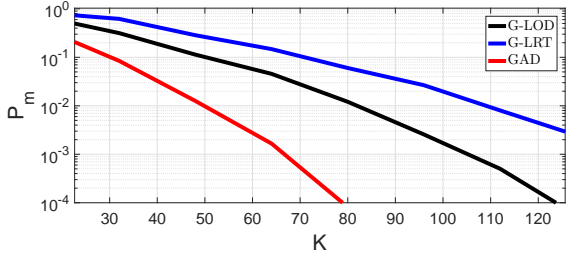}\vspace*{-2 mm}
	\caption{Global probability of missed detection versus the number of NSs $K$ for a given false alarm probability of $\Pf = 10^{-3}$, $N=10^7$, and $\mu = 2\times 10^9$.}
\vspace*{-2 mm}
\label{Fig::4}
\end{figure}

In the following, in Figs. \ref{Fig::3} and \ref{Fig::4}, we compare the performance of all proposed detectors in terms of the global false alarm and missed detection probabilities by averaging over $4\times 10^6$ independent Monte Carlo simulations, where the distributions of the received signals at the NSs and the FC are obtained based on the expressions given in (\ref{eq:local-dist}) and (\ref{eq:local-dist2}). 
The locations of the target and the FC are given in Table \ref{table:2}.
For the NSs, in Fig. \ref{Fig::3}, for each simulation, we uniformly distribute the centers of $K=64$ NSs in a 2-D square surveillance area with an edge length of $25~\mu$m and centered at the origin. To evaluate the performance of the sub-optimal detectors, we need to find the ML estimates of ${\mathbf{x}}_T$ and ${\mu}$, which may have no analytical solutions.  
Therefore, as mentioned in Section \ref{sec:GLRT}, we propose to perform a grid search as an approximation method to obtain $\widehat{\mathbf{x}}_T$ and $\widehat{\mu}$. To determine $\widehat{\mathbf{x}}_T$ and $\widehat{\mu}$, we use the grid points defined by sets $\{ (x_j, y_{j'}, 0): x_j=-12.5+25j/15, y_{j'}=-12.5+25j'/15; j,j'=0,1, ..., 15 \}$ and $\{ 2l\mu/100;, l=0,1, ..., 100 \}$, respectively, where $x_j,y_{j'}$ are in $\mu$m. Therefore, for each realization, for the G-LRT in (\ref{GLR}) and the G-LOD in (\ref{GLODF}), the maximization is reduced to a search among $101\times16^2$ and $16^2$ candidates, respectively. Note that the FC can be a complex node, e.g., a processor that can perform the grid search or a node that is connected to a computer that can perform the grid search.

In Fig. \ref{Fig::3}, we compare the performance of the proposed detectors by plotting the global probability of missed detection, $\Pm$, versus the global probability of false alarm, $\Pf$, for $N=10^7$ and $\mu \in \{6\times 10^{8}, 4\times 10^{9}\}$. 
%
%
Fig. \ref{Fig::3} shows that for $\mu = 4 \times 10^{9}$, the performance of the G-LOD is better than that of the G-LRT, while for $\mu = 6\times 10^{8}$, the G-LRT outperforms the G-LOD. 
The difference between the performance of the G-LOD and the G-LRT can be justified as follows. 
%
Since we have used a fixed number of grid points for finding $\widehat{\mu}$,
the accuracy of $\widehat{\mu}$ is worse for larger $\mu$. Therefore, we can expect that for smaller $\mu$ the G-LRT outperforms the G-LOD, as in Fig. \ref{Fig::3}, where the G-LRT performs better than the G-LOD for $\mu = 6 \times 10^{8}$.
%

In Fig. \ref{Fig::4}, we study the impact of the number of NSs $K$ on the performance of the G-LOD and the G-LRT. To this end, we show the global probability of missed detection versus $K$, given a fixed global probability of false alarm of $\Pf = 10^{-3}$, $N=10^7$, and $\mu = 2\times 10^9$. Fig. \ref{Fig::4} reveals that as the number of NSs increases, the performance of both G-LOD and G-LRT improves, since both detectors are able to exploit the independent signals received from the different NSs for performance improvement. Finally, we observe that there is a considerable gap between the (idealistic) GAD and the proposed sub-optimal detectors which suggests that the design of improved decision rules for the FC is a promising topic for future research.

\section{Conclusions and Future Work} \label{sec:conclusion}
In this paper, we studied the problem of target detection in MC systems by developing a composite hypothesis testing framework, where we assumed that the location of the target and the biomarker secretion rate were unknown at the NSs and the FC. We derived a closed-form expression for the PMF of the received signal at each NS. We then proposed a simple detector for the NSs and showed that it is UMP. Finally, we derived two sub-optimal detectors for the FC to obtain the final decision regarding the presence of a target and evaluated their performance via simulations. 

A promising topic for future work is the investigation of detection schemes for further relaxed assumptions regarding the available a priori knowledge. In particular, the case where the FC also does not know the locations of the NSs is relevant.

\end{document}